\documentclass[a4paper,10pt,twocolumn]{article}
\usepackage[utf8]{inputenc}
\usepackage{multicol,graphicx}
\usepackage{mathptmx}
\usepackage[T1]{fontenc}
\usepackage{textcomp}
\usepackage{url}
\usepackage{amsmath}
\usepackage{amsfonts}
\usepackage{amssymb}

\usepackage[T1]{fontenc}
\usepackage[english]{babel}
\usepackage{hyperref}
\usepackage[natbib=true, backend=bibtex,style=ieee,doi=false,isbn=false,url=true,maxnames=6,citestyle=numeric-comp,giveninits=true]{biblatex}

\usepackage[super]{nth}

\fontfamily{ptm}\selectfont

\usepackage{graphicx} %
\usepackage{subcaption}

\usepackage{amsmath}
\usepackage{mathtools}
\usepackage{amssymb} %

\bibliography{refs.bib} %

\usepackage{subfiles} %

\usepackage[font=small]{caption}
\usepackage{amsmath}

\usepackage{placeins}

\newcommand{\mysection}[1]{\vspace{0.4cm} \uppercase{#1} \vspace{0.4cm}}
\newcommand{\mysubsection}[1]{\hspace{10pt}\textit{#1:}}

\begin{document}
	
\setlength{\textfloatsep}{10pt plus 1.0pt minus 2.0pt}	
\setlength{\columnsep}{1cm}

\twocolumn[%
\begin{@twocolumnfalse}
\begin{center}
	{\fontsize{14}{18}\selectfont
        \textbf{\uppercase{Transferring BCI models from calibration to control: Observing shifts in EEG features}}\\}
    \begin{large}
        \vspace{0.6cm}
        Ivo Pascal de Jong, Lüke Luna van den Wittenboer, Matias Valdenegro-Toro, Andreea Ioana Sburlea\\
        \vspace{0.6cm}
        Department of Artificial Intelligence, Bernoulli Institute, University of Groningen, The Netherlands\\
        \vspace{0.5cm}
        E-mail: ivo.de.jong@rug.nl
        \vspace{0.4cm}
    \end{large}
\end{center}	
\end{@twocolumnfalse}%
]%

ABSTRACT:
Public Motor Imagery-based brain-computer interface (BCI) datasets are being used to develop increasingly good classifiers. However, they usually follow discrete paradigms where participants perform Motor Imagery at regularly timed intervals. It is often unclear what changes may happen in the EEG patterns when users attempt to perform a control task with such a BCI. This may lead to generalisation errors. We demonstrate a new paradigm containing a standard calibration session and a novel BCI control session based on EMG. This allows us to observe similarities in sensorimotor rhythms, and observe the additional preparation effects introduced by the control paradigm. In the Movement Related Cortical Potentials we found large differences between the calibration and control sessions. We demonstrate a CSP-based Machine Learning model trained on the calibration data that can make surprisingly good predictions on the BCI-controlled driving data.

\mysection{Introduction}

The public availability of various BCI datasets has allowed for more transparent and more reliable progress in the development of Machine Learning models for EEG processing. %
The most convenient datasets to collect and make Machine Learning models for assume cue-based BCIs. These have cleanly separated instances of the various classes, which increases consistency and makes for a clear classification task.

However, such Machine Learning benchmarks often do not align with the EEG processing that a BCI with high usability needs. BCI competition IV \cite{tangermann2012review} dataset 1 \cite{blankertz2007non} attempts to address this by aiming for Motor Imagery classifiers where the cue is not known in the EEG processing. This dataset has a training section with visual cues for left hand, right hand and foot Motor Imagery, and a test session with auditory cues for left hand, right hand, foot and \textit{rest} conditions. The candidate models then need to predict for all instances in the test session which of the four states (including the \textit{rest}) the subject is in. 

This dataset bridges a gap from the classical trial-based EEG classification to BCI systems that need to make predictions without trial information. However, it also has some limitations that we aim to address with the introduction of a newly collected dataset. 

\cite{blankertz2007non} highlights that the subjects will have a transient phase between hearing the auditory cue and performing the corresponding task. The timing of this is not precisely known, so the models are not evaluated on these transient phases. As a result, these models might have unexpected behaviour on these transient phases, and will not be optimised to detect the onset of a new state. This can be a problem when BCIs require low latency, which is important for learning to use a BCI \cite{xu2013enhanced}.

Like most paradigms, it also does not give a good reflection of the mental state of a user using the BCI to achieve a task. The transition from a BCI model that works well in a controlled paradigm to using the BCI to perform a control task introduces many unknowns. The EEG patterns may change due to planning, eye movements or visual attention and it is generally unclear exactly what does and does not change when shifting from a calibration paradigm to a control task.

\mysubsection{Contributions}
To address these issues and offer a dataset that is better suited to transfer a BCI-based ML model from calibration to control, we demonstrate a new paradigm with a preliminary data analysis. The paradigm has a visual cued calibration session similar to BCI competition IV dataset 1 \cite{blankertz2007non}. However, the testing session has the subject drive a simulated car. The steering of the car is done through the detection of the flexion of the left and right hand based on electromyography (EMG) signals. By using the EMG to control the car we can observe the EEG of a subject as if they are using a BCI for a control task. The task is then to predict the motor execution state (as measured by the EMG), based on the EEG patterns from the motor cortex.

\begin{figure}[h!]
    \centering
    \includegraphics[width=\linewidth]{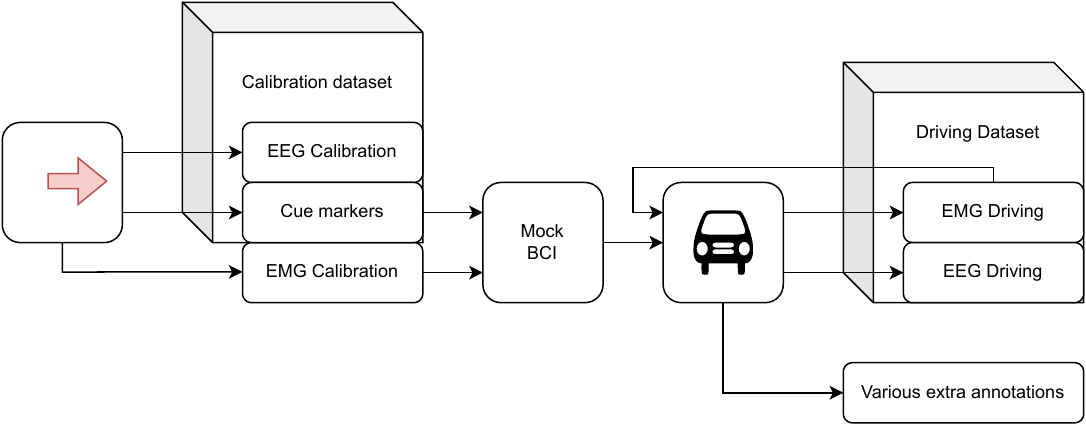}
    \caption{Design of the data acquisition. First subjects do a calibration session following the Graz-BCI Motor Imagery paradigm. The EMG for this is used to develop a \textit{mock BCI}, which the users then use to control a simulated car. The EMG, EEG and markers are recorded in both sessions resulting in a driving dataset and a calibration dataset.}
    \label{fig:motorbci_pipeline}
\end{figure}

The track is designed to have left turns, right turns, and straight sections that can be used as artificially segmented trials. The EMG can then be used to determine the motor execution onset, and allow us to investigate differences in EEG patterns between the calibration session and the driving session.

We show a preliminary analysis of a dataset recorded under this paradigm looking at Sensorimotor Rhythms (SMR) and Movement Related Cortical Potentials (MRCP) for calibration and compare these to the artificially segmented trials during driving. We also demonstrate classification with multiclass CSP \cite{grosse2008multiclass} in the calibration session, the driving session, as well as a classifier trained on the calibration and applied to the driving session. 

We believe that the introduction of this paradigm will allow for the experimentation with motor decoding models that are exceptionally well suited for transferring from the benchmark to the user.

\mysection{Methods}

The implementation of the paradigm focuses on building a \textit{mock} BCI that the subjects can use to perform a control task. The goal is to collect all the EEG data as if the subjects are using a BCI, without the risk of complete loss of control introduced by using a real EEG-based BCI. 

The design for the current study consists first of a calibration session following the Graz-BCI Motor Imagery paradigm \cite{pfurtscheller2003graz}. The EMG from the calibration session is used to make an EMG-based classifier, which will function as our mock BCI. The EEG is recorded for offline analysis. After this, the EMG-based mock BCI is used to make online predictions during a simulated driving task. This gives us the EEG patterns of our test subjects using the mock BCI, with EMG as the ground truth of their control intentions.\footnote{The driving paradigm is available at \url{https://github.com/lukeluna/continuous_control}, the EMG classifier at \url{https://github.com/ivopascal/emg_classifier} and the analysis code at \url{https://github.com/ivopascal/continuous_control_bci}. The recordings will be made available in the camera-ready paper.} The whole setup of the recording is visualised in Figure \ref{fig:motorbci_pipeline}.

In an offline analysis, we can then look at the EEG during calibration using the cues as the ground truth and the EEG during driving using the EMG as the ground truth. Within the calibration and driving sessions we can look at the MRCP and SMR, and we can develop classifiers on the calibration EEG and apply them to make predictions with the driving EEG. %

This study was conducted with 20 healthy subjects aged 19-45  years old ($\mu=26, \sigma^2=40$), although the data from one subject was removed from the analysis because a section of the driving was not recorded. None of the subjects had prior BCI experience. The EMG, EOG and EEG were recorded using the Biosemi ActiveTwo. 32 EEG channels were collected following the 10-20 system. Two monopolar EMG channels were measured on each forearm to correspond with wrist flexion and wrist extension, resulting in four total EMG channels. Four EOG channels were measured to capture horizontal and vertical eye movements. All 40 channels were recorded at 2048Hz. Subjects sat in a chair facing a computer with their arms resting on a desk at a comfortable height. A towel was placed under the forearms to elevate the wrists slightly which allowed for easy wrist flexion. 

Whenever subjects needed to perform a "left" command they flexed their left hand inward. The "right" command corresponded with flexing the right hand inward. At the end of each trial or turn they returned their hand to a forward resting position.

\begin{figure}
    \centering
    \includegraphics[width=\linewidth]{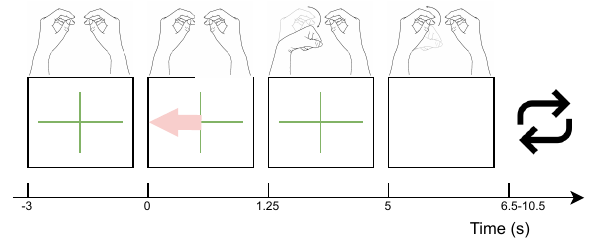}
    \caption{Timing of the calibration paradigm from OpenVibe. %
    }
    \label{fig:timing_diagram}
\end{figure}

\mysubsection{Calibration Paradigm} \label{sec:training}
The Motor Imagery paradigm from OpenVibe\cite{renard2010openvibe} version 3.5.0 was used without any modification to the timing. This entails a 30 second preparation time, followed by 2x20 shuffled trials of left and right hand motor execution. Each trial starts with a cross displayed for 3 seconds as a preparation cue. Then an arrow pointing left or right is displayed for 1.25 seconds. When this disappears the subject performs the movement and keeps the wrist flexed while looking at the cross. After 3.75 seconds this cross disappears and the subject moves their hand back to a resting position. A random rest period of 1.5 to 3.5 seconds separates each of these trials. The timing of the paradigm is visualised in Figure \ref{fig:timing_diagram}. We set $t=0$s to the moment that the participant knows the direction, so the movement onset is at $t=1.25$s

We chose to have the movement performed after the arrow disappears, instead of when it appears. This way, the participant is able to prepare the action belonging to the visual cue and initiate the movement when the arrow appears. This should provide a more consistent movement onset, and allows effects of movement planning and inhibition to be included in the calibration data.

In between the left and right hand trials 40 rest samples were extracted, from 0.5s after the end of the previous trial, until 4.5s after. This means it is partly recorded while the subject is looking at a blank screen, and partly when the fixation cross of the next trial is shown. This makes the rest slightly more noisy, but also more similar to rest periods during the driving session. These trials will be used for the online EMG classifier, as well as for the offline analysis of the calibration dataset.

\mysubsection{EMG Classifier} We train a subject-specific online EMG classifier on the EMG collected during the calibration session. This EMG classifier will be used during the driving session to control the simulated car.

The EMG is measured with 4 monopolar electrodes to measure two muscles on each arm. Specifically the flexor carpi radialis and the extensor carpi radialis longus.

First, the EMG of these 4 channels is re-referenced using Common Average Reference (CAR) over the EMG channels. Then the EMG is band-pass filtered between 30-500Hz and notch filtered at 50Hz using causal 
\nth{4} order Butterworth filters. Since the online classification should have low latency, the trials are cropped to the middle 200ms of each movement from $t=6.125$s to $t=6.145$s. We found this to have a minimal impact on classification performance. The mean power of the four EMG channels are used as four features for an LDA classifier. Using 10-fold cross-validation we find that this per-subject EMG classifier has a mean accuracy of 94\% with a standard deviation of 5\%.

The online implementation of the EMG classification uses Lab Streaming Layer (LSL) \cite{kothe2014lab}. LSL allows for real-time streaming of the EMG recordings from BioSemi to the EMG classifier (implemented with MNE-Python \cite{GramfortEtAl2013a}), and streaming the classifications to the driving environment. It also saves the multiple streams and provides precise alignment of timestamps from the various streams.

\mysubsection{Driving Paradigm} \label{sec:testing_paradigm}
\begin{figure}
    \centering
    \includegraphics[width=\linewidth]{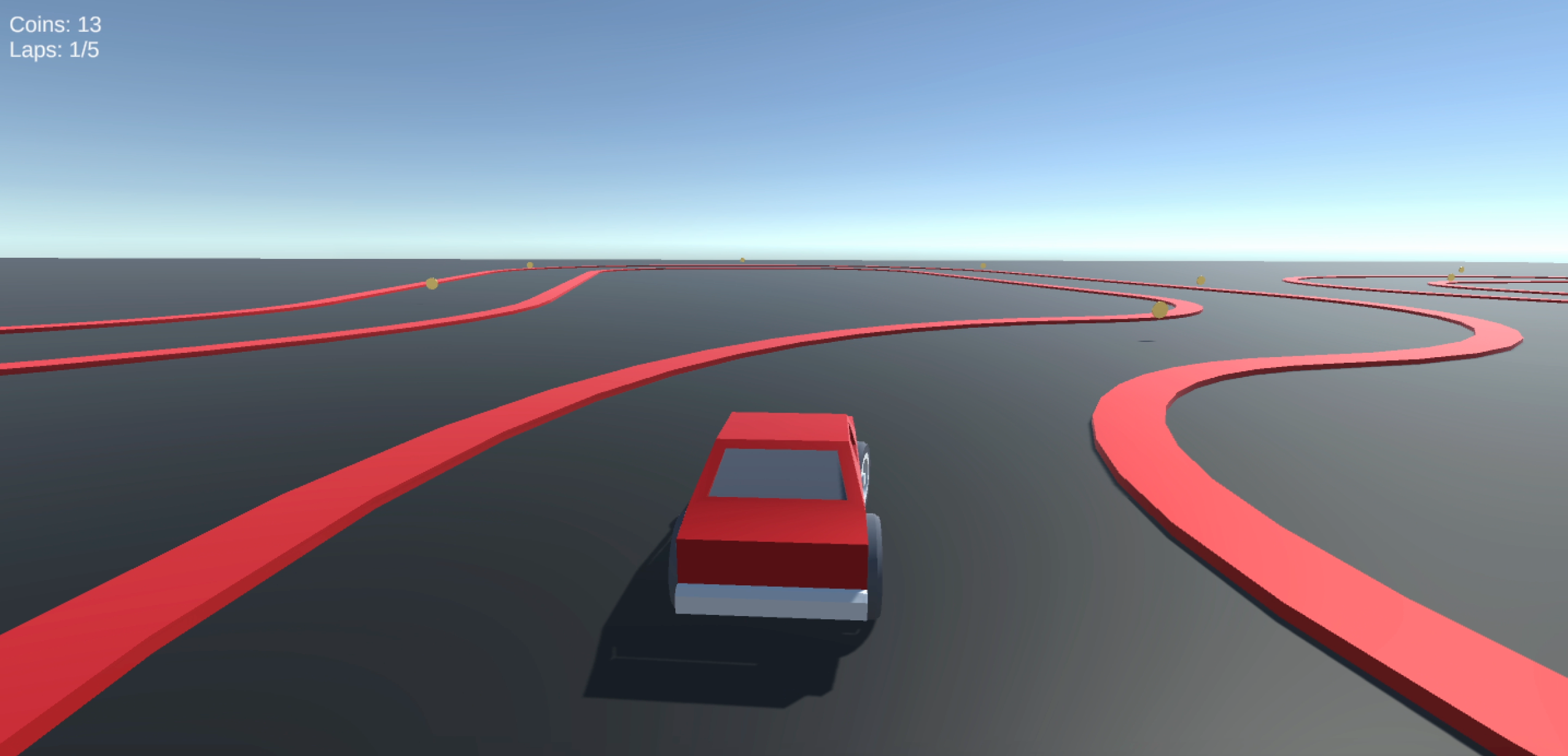}
    \caption{Participant view during the driving session.}
    \label{fig:driving_3d}
\end{figure}
As a control task, we chose to control a simulated car in Unity. At the start the car accelerates until it reaches a predefined constant speed. The speed is kept consistent during the driving so the participant needs to initiate and terminate turns at the right time. We believe this is an important factor in developing a BCI for control tasks, as early preparation of a movement may cause patterns in the premotor cortex which may be confused with the actual movement. The ability to distinguish between movement preparation and movement execution is necessary to be able to have a good estimate of the movement onset.

The 3D driving environment is shown in Figure \ref{fig:driving_3d}. The choice of a 3D design allows for better engagement, but may also affect the EEG with eye movements, visual perception effects, or planning effects. This makes the recorded EEG more ecologically valid for real BCI control. 

Each participant drives around the track for 5 laps. After these 5 laps, they have a break and start again when they feel ready. They perform this task 3 times for a total of 15 laps, resulting in an average of 37.5 minutes ($\pm 6$ minutes) of driving per participant. 

From the driving session, we extract trials of left turns, right turns, and straight sections by identifying periods of at least 3.75 seconds where the EMG classifier made the same prediction. This allows us to investigate the SMR and MRCP for left and right turns, and gives rest trials sufficiently similar to the calibration session. Any trials with peak-to-peak differences exceeding $100\mu V$ after epoching are rejected.

\mysection{Results}

\mysubsection{Sensorimotor Rhythms} \label{sec:SMR}
We look for the Event Related Desynchronisation (ERD) that is commonly found in Motor Imagery, Motor Attempts \cite{chen2021differences} and Motor Execution \cite{pfurtscheller2000spatiotemporal}. Specifically, we look for contralateral ERD in the $\alpha$ band around $8-12$Hz, which may start before movement onset due to movement planning. We may also find an ERD in the $\beta$ band around $12.5 - 30$Hz.

We apply CAR followed by a non-causal FIR band-pass filter in the range $[1, 35]$Hz with a lower transition bandwidth of $1$Hz, an upper transition bandwidth of $8.75$Hz and a filter length of 3.3s. Then we apply artefact removal with FastICA \cite{hyvarinen2000independent}, and take the Surface Laplacian \cite{kayser2015benefits} around channels C3 and C4. This gives us the relevant frequencies around the parts of the motor cortex responsible for left hand and right hand movement while minimising artefacts. We take epochs from both calibration and driving sessions such that the movement starts at $t=1.25$. Relative to the movement onset we look at the time-frequency effects from $t=-3$ to $t=5$. This gives us an indication of the activation before movement onset, and for the remainder of the trial.  

We then use DPSS multitapers to determine the time-frequency response from $[5, 35]$Hz, at increments of 1Hz. For this analysis, the epochs are temporarily padded with 0.5s of leading and trailing EEG data to avoid edge effects. Unlike common ERD visualisations we do not subtract the baseline activation. The baseline sections between calibration and driving are very different, which would make it difficult to distinguish changes in the baseline from ERD/ERS effects. Instead, we look at the absolute time-frequency plots. 

\begin{figure}[t]
    \centering
    \begin{subfigure}{\linewidth}
        \centering
        \includegraphics[width=\linewidth]{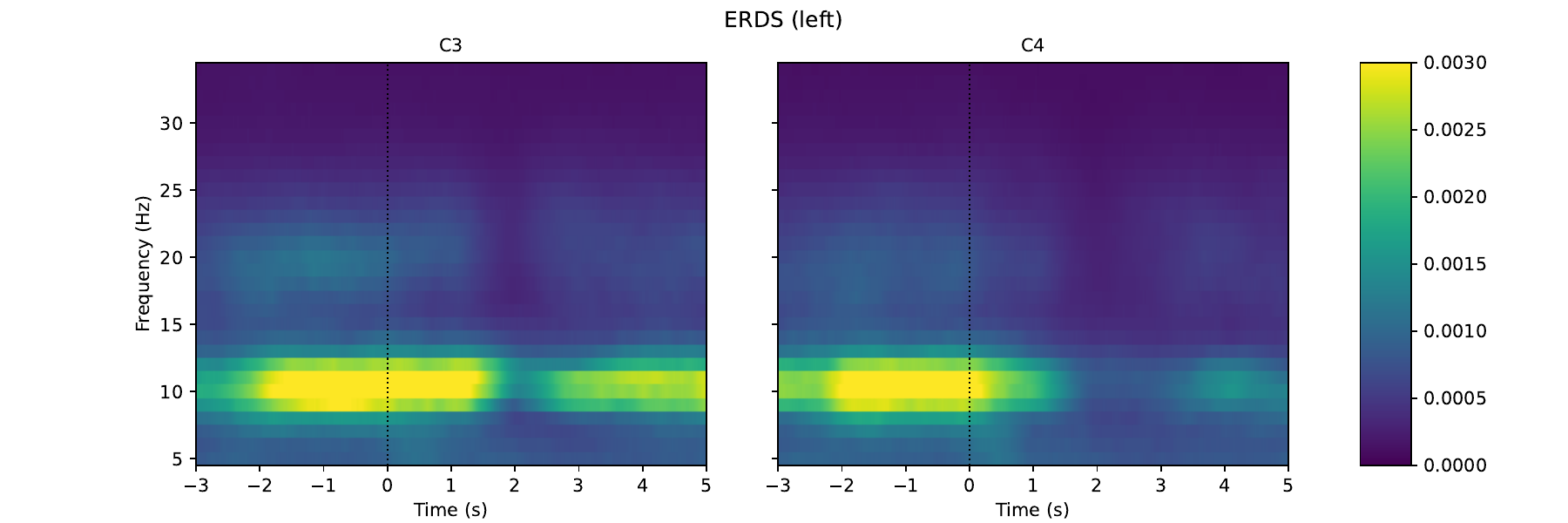}
        \caption{Left calibration}
    \end{subfigure}
    
    \begin{subfigure}{\linewidth}
        \centering
        \includegraphics[width=\linewidth]{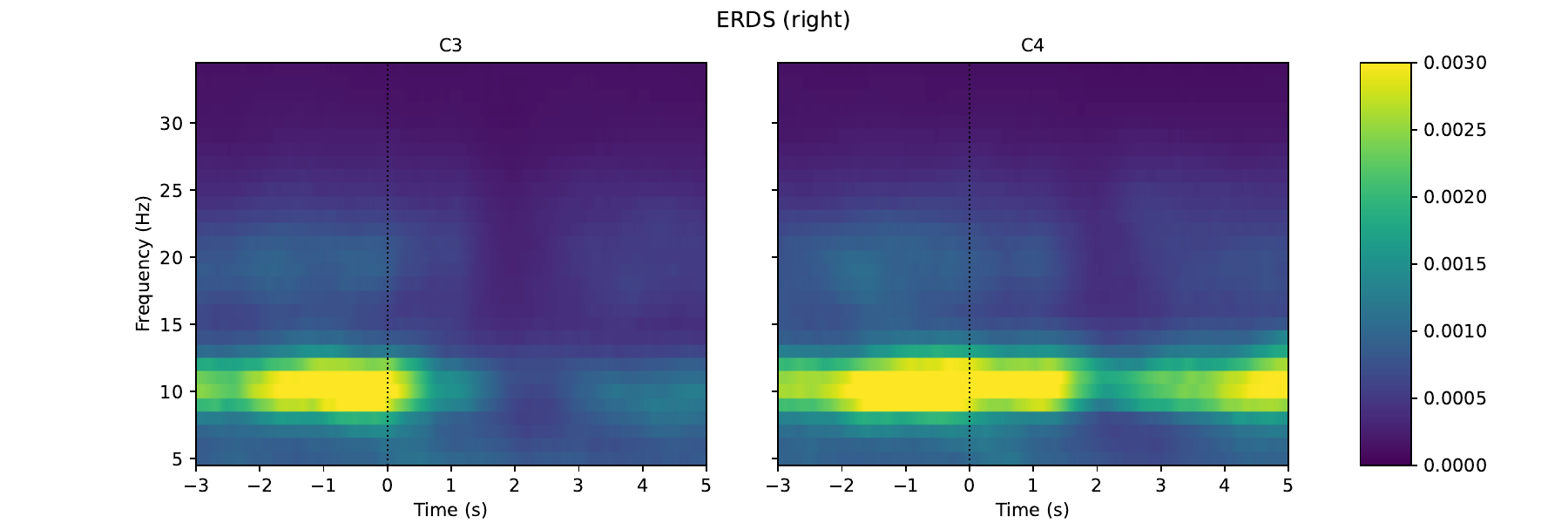}
        \caption{Right calibration}
    \end{subfigure}
    
    \begin{subfigure}{\linewidth}
        \centering
        \includegraphics[width=\linewidth]{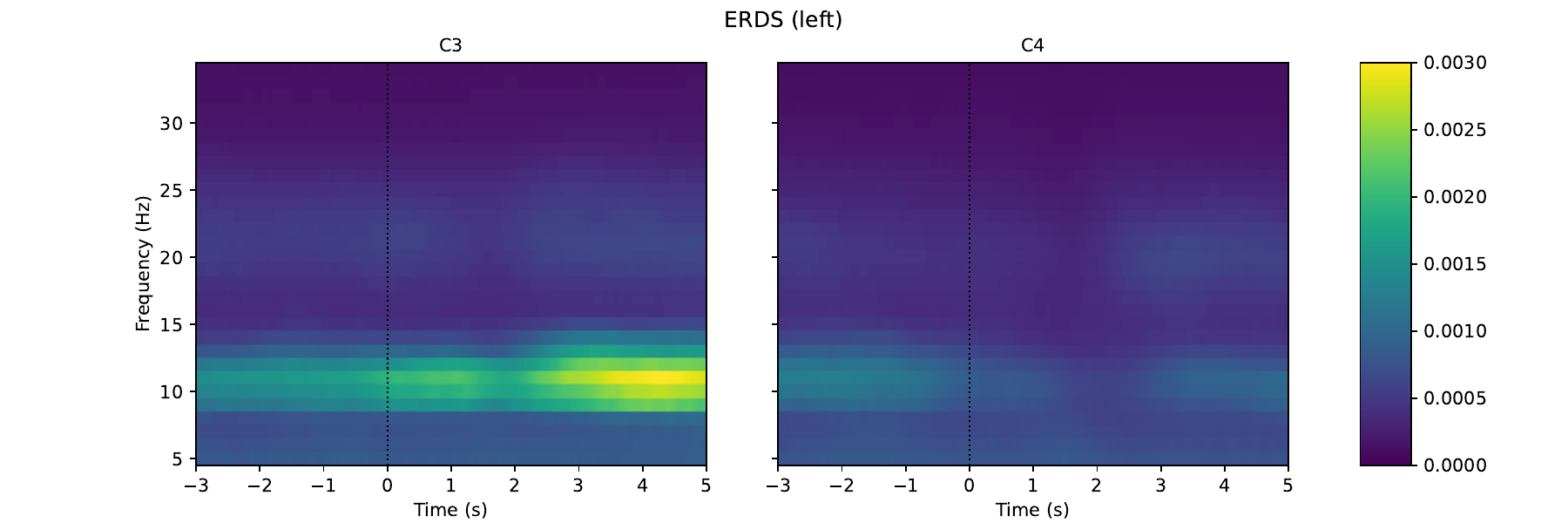}
        \caption{Left driving}
    \end{subfigure}
    
    \begin{subfigure}{\linewidth}
        \centering
        \includegraphics[width=\linewidth]{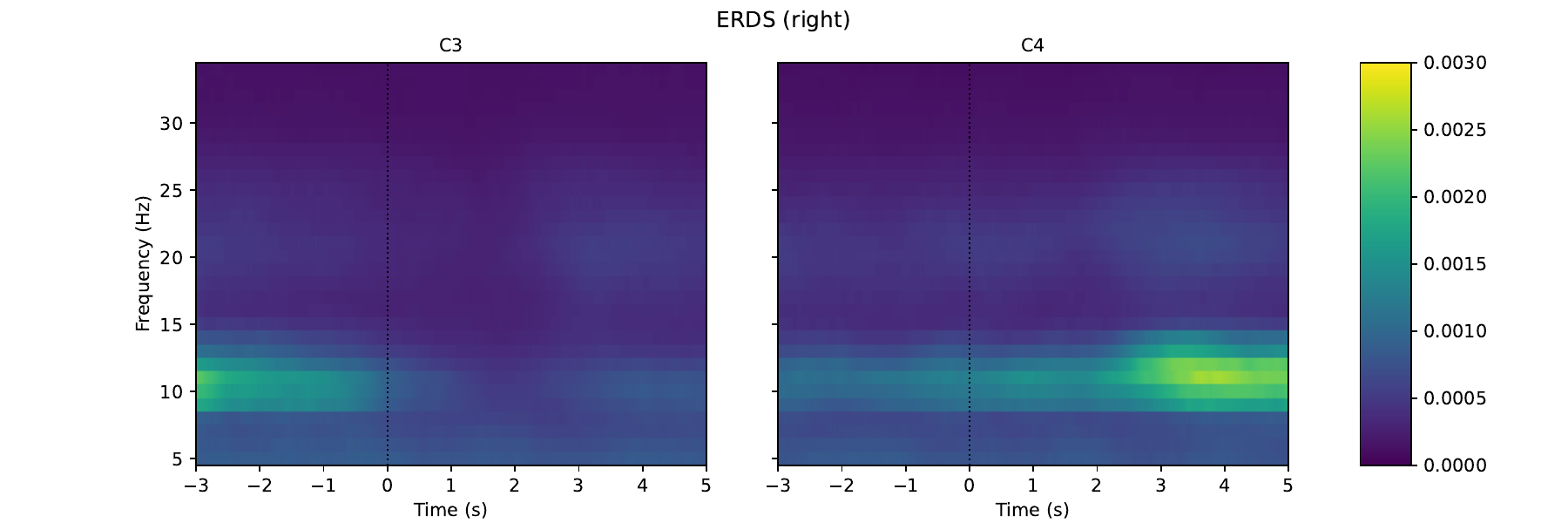}
        \caption{Right driving}
    \end{subfigure}

    \caption{SMR during the left and right trials for calibration and driving. No baseline is used. %
    }\label{fig:erds}
\end{figure}

This pipeline is applied to both left hand and right hand movements, both during calibration and driving. This allows us to see the differences that may affect decoding during calibration and driving. Figure \ref{fig:erds} shows the time-frequency response during calibration and driving for both left and right trials averaged over the population. 

In both cases, the movement onset is at $t=1.25$s. We see a contralateral decrease in the $\alpha$ band for both cases, starting slightly before movement onset. In the calibration session, the decrease starts to appear around $t=1$s, roughly 0.25 seconds before movement onset. In the driving session, the $\alpha$ band decrease starts as early as $t=0$s. Around $t=4$s we see a slight rebound in the $\alpha$ band for calibration. This rebound also appears in the driving session starting at $t=3$. This early rebound may be because the hand is in a consistent flexed position, so there is no more hand movement. 

\captionsetup[subfigure]{justification=centering}

In all figures we see a short dip in the $\beta$ band in both channels, aligned with the onset of the movement. The timing of the dip in the $\beta$ band may allow it to be used to determine the onset of the movement, while the contralateral power decrease in the $\alpha$ band may be used to determine which hand was moved.

\mysubsection{Movement Related Cortical Potentials (MRCP)} \label{sec:MRCP}
We investigated MRCPs as they are commonly used to study motor preparation effects \cite{shakeel2015review}. Since MRCPs are low-frequency phenomena we apply a causal \nth{8} order butterworth band-pass filter in the range $[0.1, 3.0]$Hz. After this artefacts are removed with FastICA and the trials are then epoched so that the movement onset is at $t=1.25$s. 

\begin{figure}
    \centering
    \begin{subfigure}{0.5\linewidth}
        \centering
        \includegraphics[width=\linewidth]{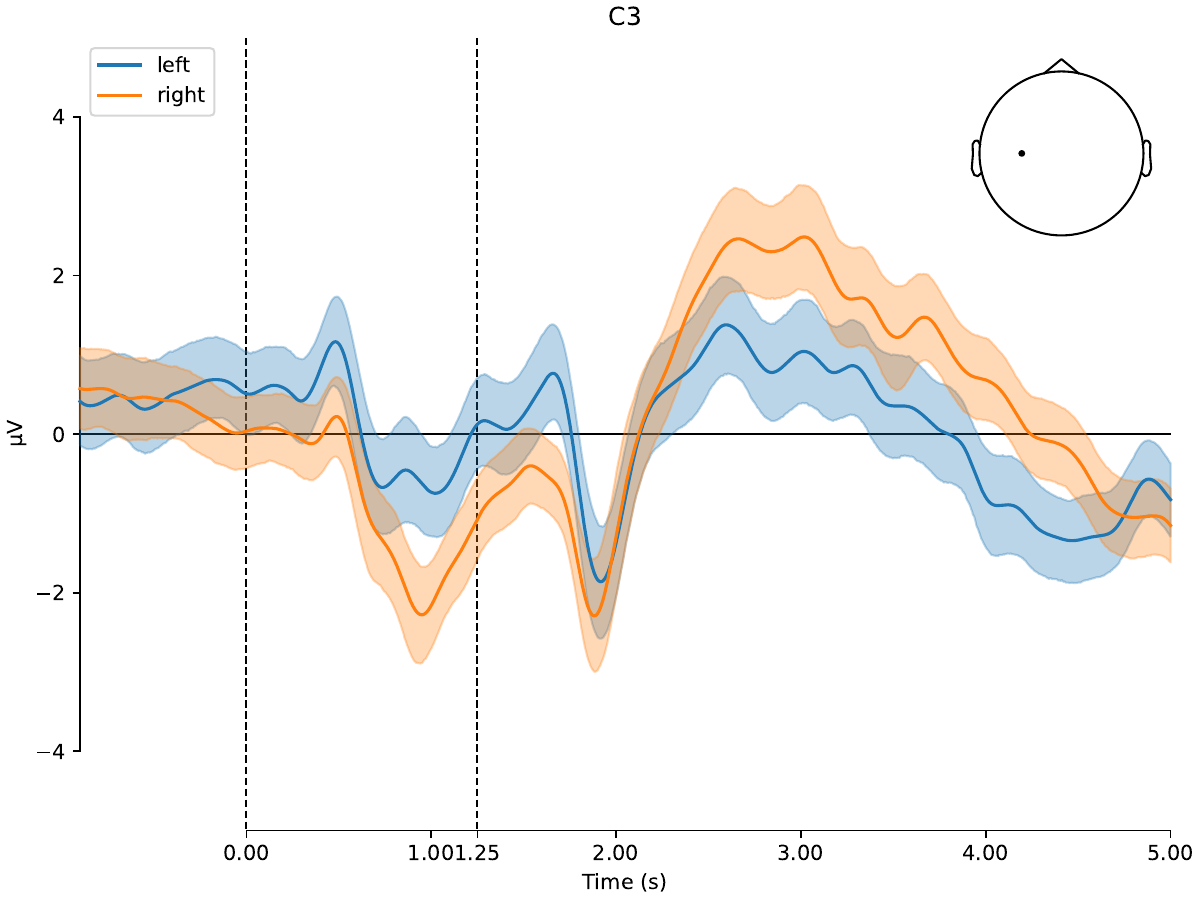}
        \caption{C3 calibration}
    \end{subfigure}%
    \begin{subfigure}{0.5\linewidth}
        \centering
        \includegraphics[width=\linewidth]{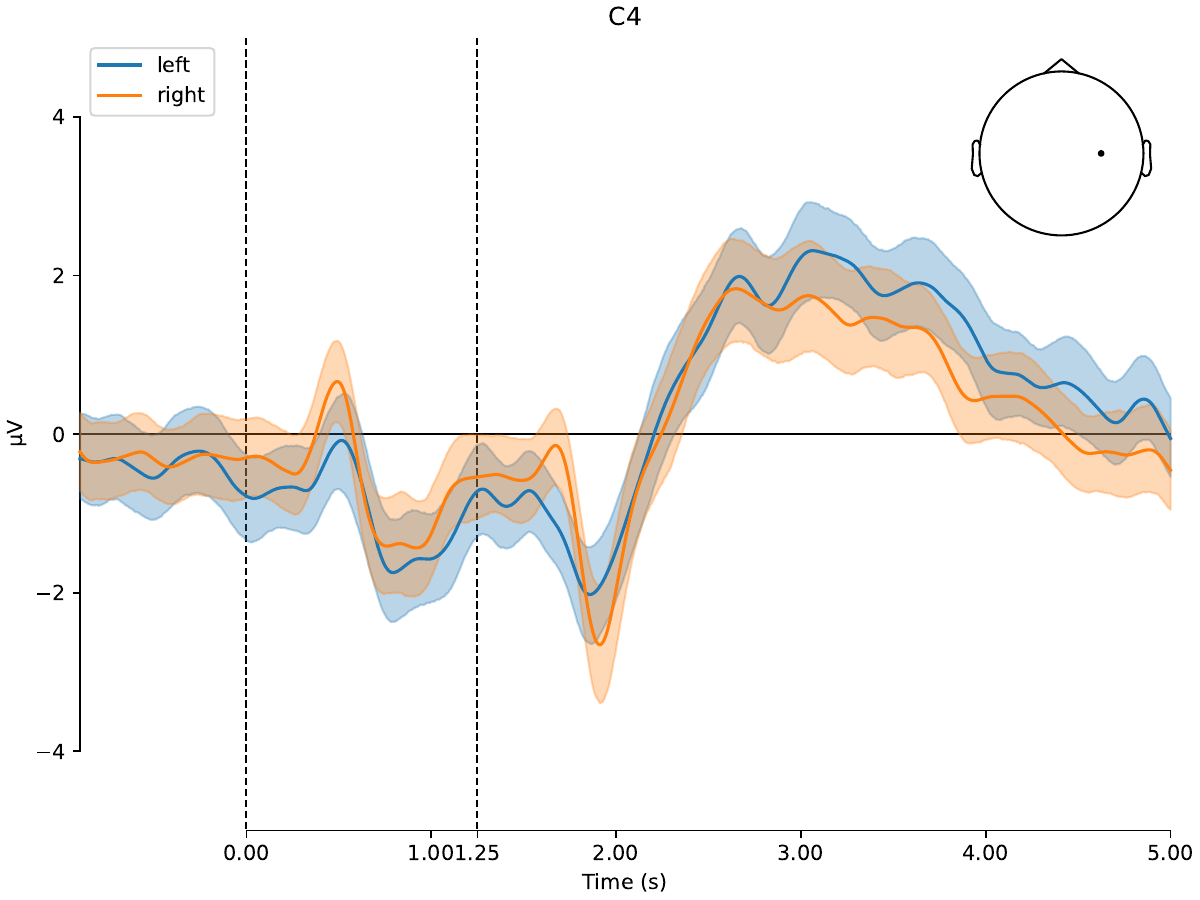}
        \caption{C4 calibration}
    \end{subfigure}
    
    \begin{subfigure}{0.5\linewidth}
        \centering
        \includegraphics[width=\linewidth]{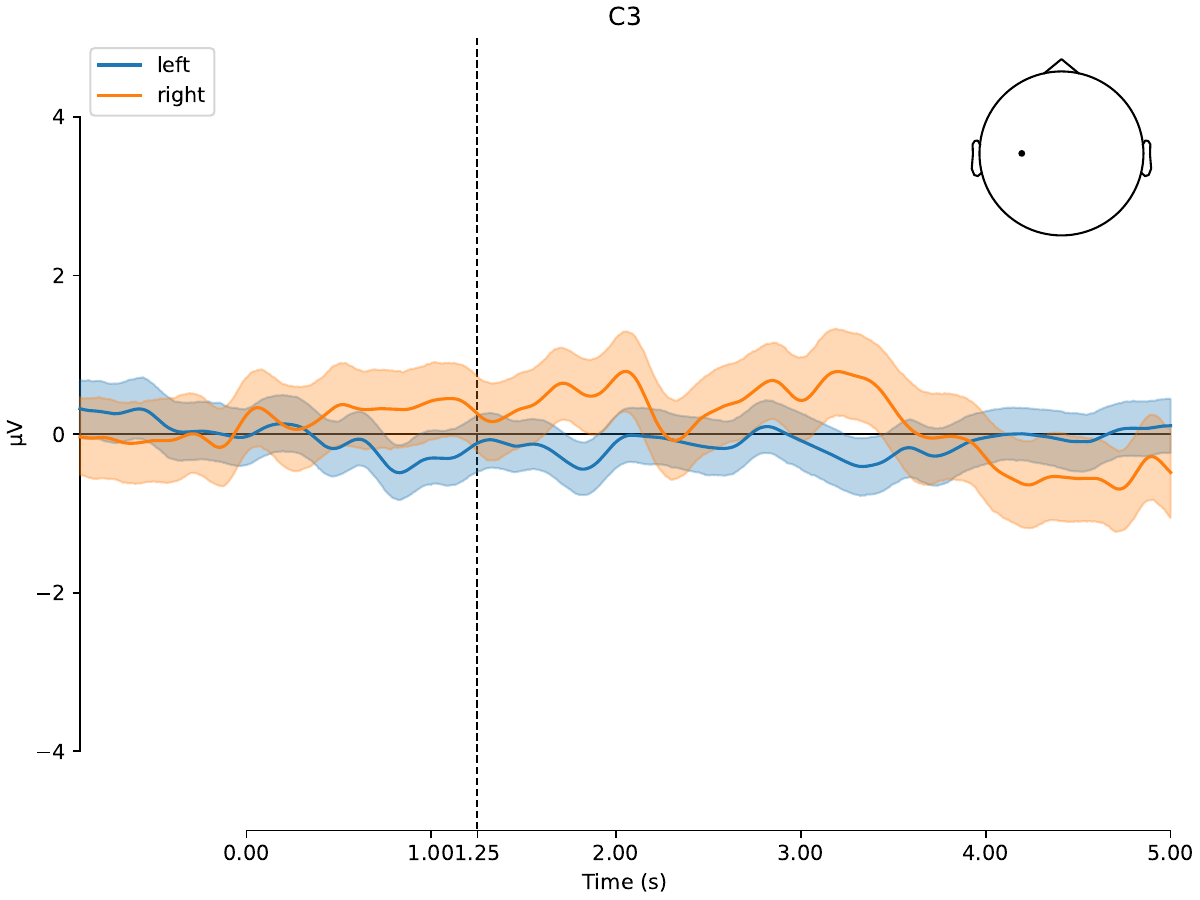}
        \caption{C3 driving}
    \end{subfigure}%
    \begin{subfigure}{0.5\linewidth}
        \centering
        \includegraphics[width=\linewidth]{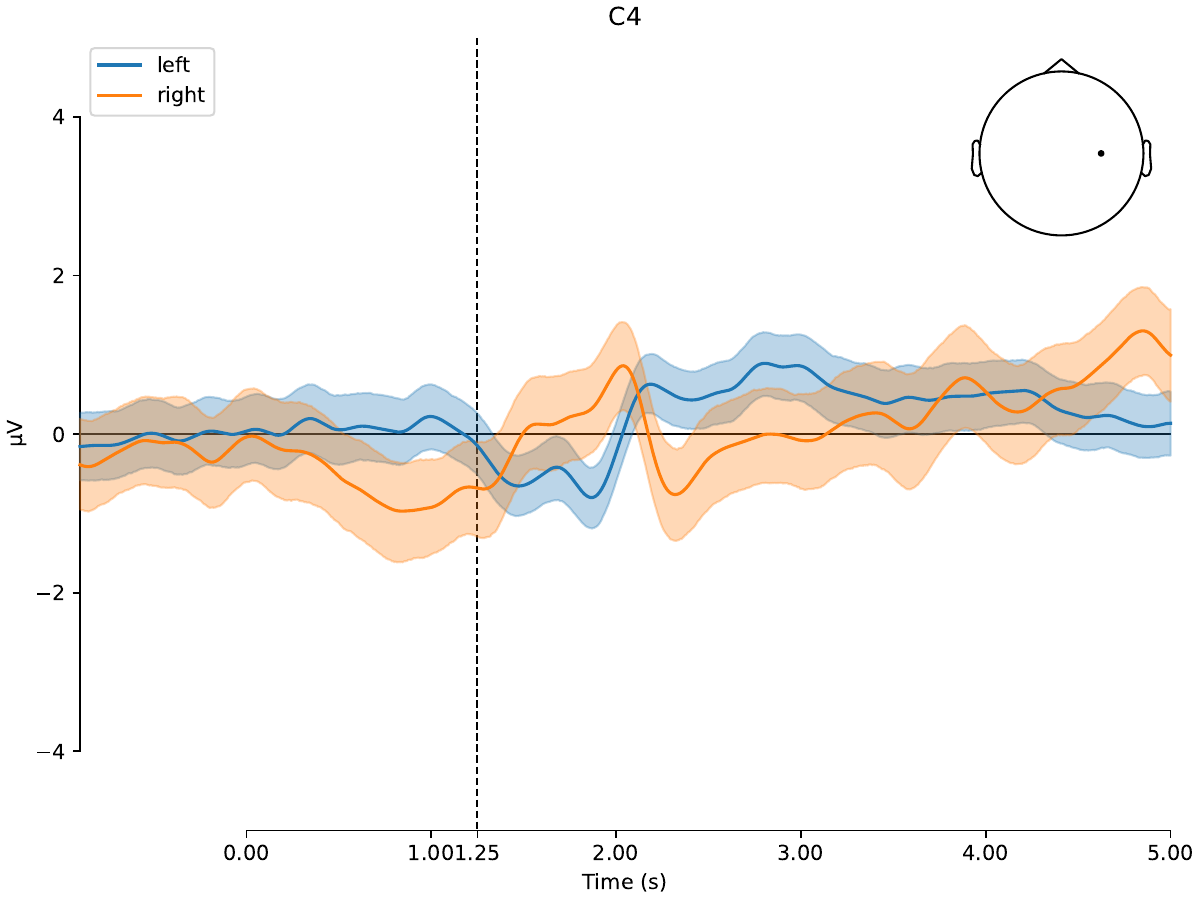}
        \caption{C4 driving}
    \end{subfigure}

    \caption{Population averaged MRCP during calibration and driving. The movement starts at $t=1.25$, indicated by the rightmost dashed line. In the calibration plots the dashed line at $t=0$ shows when participants are given the directional cue. The blue line indicates the average ERP for left trials, the orange for right trials. %
    } \label{fig:mrcp}
\end{figure}

Figure \ref{fig:mrcp} shows the MRCPs at C3 and C4 for calibration and driving. During the calibration, we observe a negative peak slightly before the movement onset at $t=1$s, which is the contingent negative variation \cite{walter1964contingent}. This effect disappears in the driving session. There are clear patterns that distinguish the classes during calibration, but they do not transfer well to the driving session.

\mysubsection{CSP classifier}
To demonstrate the ability to classify individual samples based on SMR patterns we implemented a simple CSP-based classifier.

\begin{figure}[b]
    \centering
    \begin{subfigure}{\linewidth}
        \centering
        \includegraphics[width=\linewidth]{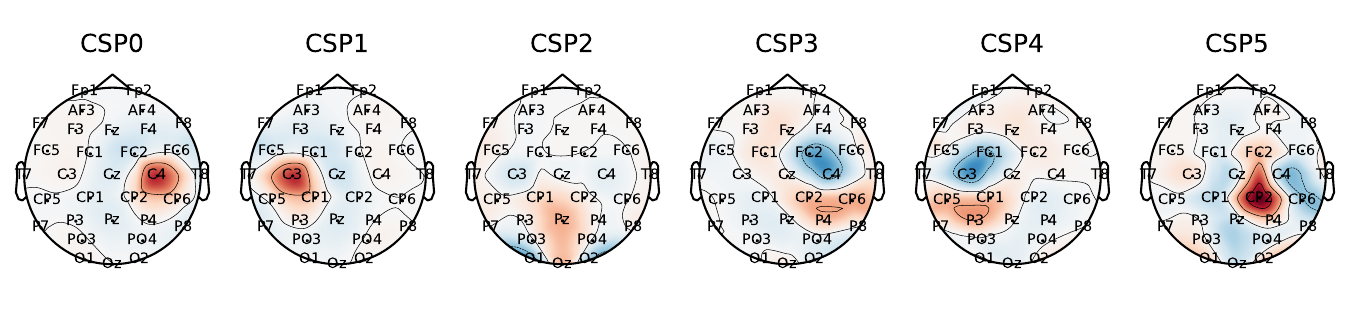}
        \caption{CSP filters calibration}
    \end{subfigure}
    
    \begin{subfigure}{\linewidth}
        \centering
        \includegraphics[width=\linewidth]{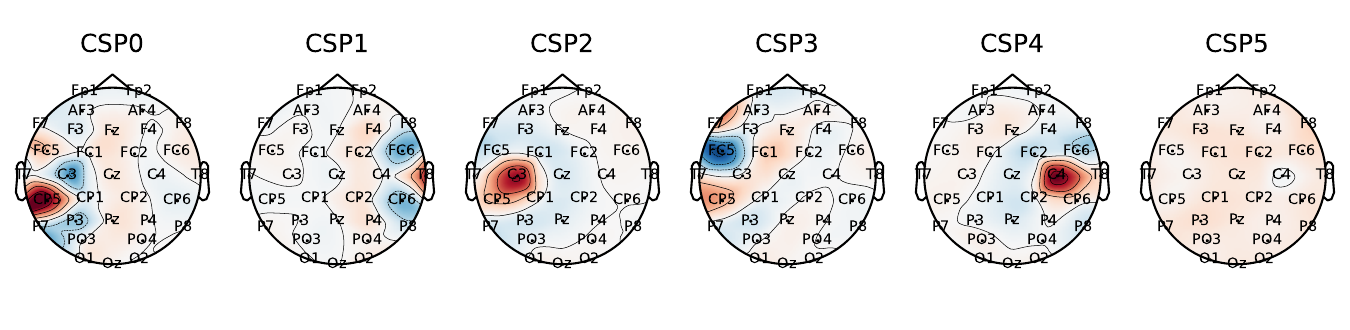}
        \caption{CSP filters driving}
    \end{subfigure}

    \caption{CSP filters for driving and calibration over the whole population. %
    The CSP filters are ordered according to Mutual Information.}\label{fig:csp_filters}
\end{figure}

First, the EEG data is re-referenced with CAR and the data is filtered with a non-causal FIR band-pass filter in the range $[1, 35]$Hz with a lower transition bandwidth of $1$Hz, an upper transition bandwidth of $8.75$Hz and a filter length of 3.3s. Then, FastICA is used to remove artefacts. The ICs were fitted and rejected or kept based on the calibration data, but used on both the calibration and the driving data. This makes the process suitable for online implementations. Epochs are taken from movement onset $t=1.25$s to the end of the trial $t=5$s. Rest samples are also used from between the trials in calibration, or from straight sections in driving. 

Six multiclass CSP filters \cite{grosse2008multiclass} regularised with shrinkage are used to extract the features. The logarithm of the band power for each filter is used by a Logistic Regression classifier to predict the classes. This pipeline is fitted 5 times with 5-fold cross-validation to make the predictions. The model is then trained once more on all the data to show the CSP filters. 

\begin{figure}
    \centering
    \includegraphics[width=\linewidth]{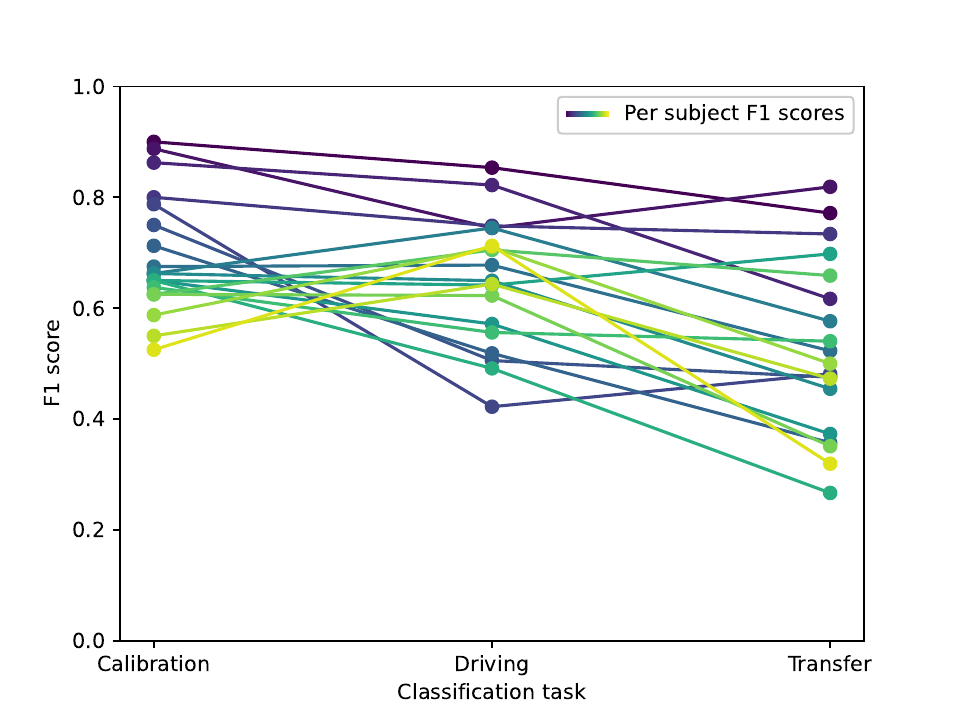}
    \caption{F1 scores of the CSP classifiers in different sessions. We show trained and tested on calibration, trained and tested on driving, and lastly, the transfer scenario is trained on calibration and tested on driving. Each line represents one subject.  %
    }\label{fig:csp_f1_scores}
\end{figure}

This procedure is applied once to the calibration data, and once to the driving data. Lastly, we also train the model on all the calibration data and use it to make predictions on the driving data. Transferring the model in this way means it is never trained on EMG data but is still able to predict the EMG data.

We use the first 10\% of driving trials to calibrate a rest threshold, addressing a shift in class imbalance between calibration and driving. The transfer performance is then evaluated on the remaining 90\% of driving trials. The threshold that would need to be used in an online scenario should be set manually. %

Figure \ref{fig:csp_filters} shows the learned CSP filters for the calibration and driving sessions for the whole population, ordered by Mutual Information. In the calibration CSP0 and CSP1 clearly identify the relevant areas of the motor cortex for left and right hand motor control. CSP3 and CSP4 show bipolar effects around the motor cortex. The CSP filters learned on the driving data are not as clean. However, CSP2 and CSP4 still correspond to the relevant parts of the motor cortex. There are no discernible effects in the prefrontal areas, indicating that the learned patterns are not due to eye movement artefacts.

Figure \ref{fig:csp_f1_scores} shows the F1 scores for each subject in each session. The performance is best in the \mbox{calibration session ($\mu= 0.69, \sigma^2=0.011$)}, followed by the driving session \mbox{($\mu=0.65, \sigma^2=0.013$)}, followed by the transfer scenario \mbox{($\mu=0.53, \sigma^2=0.025$)}. The subjects with poor performance on the calibration data are also more likely to have poor performance in transfer, but this is not consistent.

\begin{figure}
    \centering
    \includegraphics[width=\linewidth]{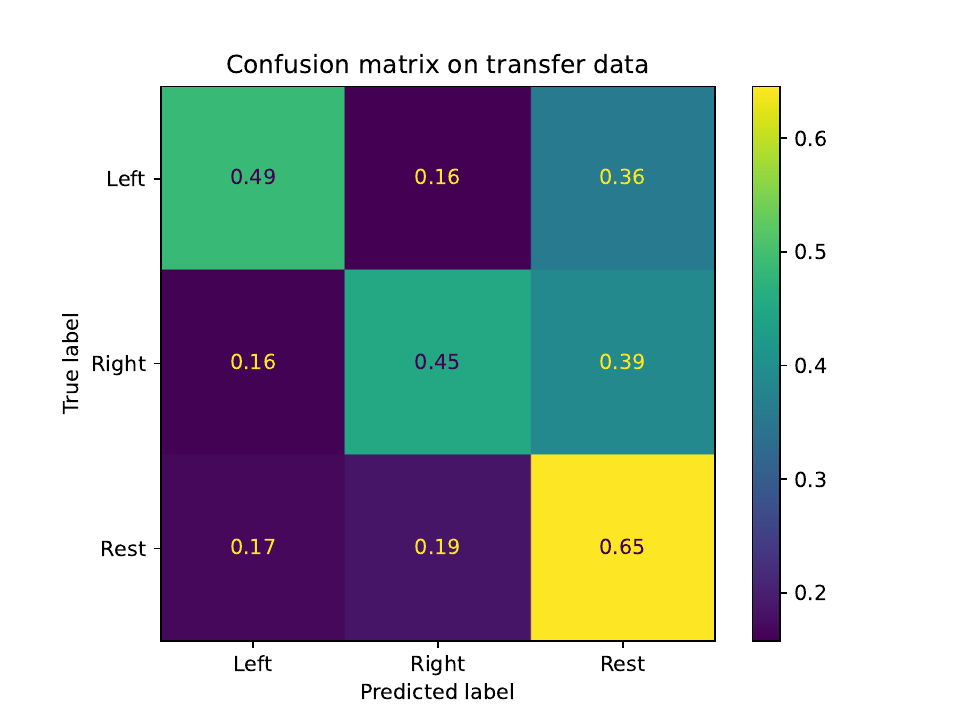}
    \caption{Population averaged confusion matrix for transfer. %
    }\label{fig:confusion-matrix}
\end{figure}

Lastly, Figure \ref{fig:confusion-matrix} shows the confusion matrix averaged over the population in the transfer scenario. A model was trained for each subject based on their calibration data, and then tested on their driving data. The (mis)classifications were then aggregated into this confusion matrix. There are few cases where a left hand is mistaken for a right hand, or the other way around, but many movement samples are still mistaken for rest samples. 

\mysection{Discussion}

The SMR analysis showed that there are patterns to observe in the calibration phase, that remain mostly consistent when people are using a BCI for a control task. We notice that there is a longer movement preparation effect and that the average $\alpha$ power is lower during the driving task. The dip in the $\beta$ band corresponds with the movement onset both in the calibration and in the driving session, and is minimally affected by preparation effects.  

We found that the MRCPs are different between calibration and driving. ERPs identified in the calibration data are not identifiable in the driving data. This may be because the two paradigms are too different, or because the 200ms EMG classification interval gives too imprecise movement onsets. 

From the MRCPs and the SMR we find that a classifier that needs to be transferable from calibration to driving is best suited using band power features. The subsequent CSP-based classifier was able to distinguish between left and right hand trials surprisingly well, but it had quite a few false positives for the rest trials. As expected, accuracy becomes worse in driving due to the added stimulation of the environment, and the accuracy becomes even worse when transferring from calibration to driving. However, both of these factors have a relatively minor impact compared to the individual differences between subjects. 

The CSP filters showed that the models are picking up phenomena from the motor cortex, indicating that the patterns are originating from movement (intention) and not from eye artefacts or visual attention. The fact that the ICA was fitted to the calibration, and used in the driving makes this design suitable for use in an online BCI.

\mysection{Conclusion}

The findings from the SMR analysis, MRCP analysis and the CSP classifier show that an SMR-based classifier trained on calibration and applied for a control task is feasible. The SMR patterns are largely similar between the calibration and the driving session, allowing for a surprisingly small decrease in classification accuracy. 

However, the current analysis still leaves some hurdles for the implementation of an online EEG-based BCI to control the driving task that can be addressed by future analyses on this dataset.

The primary limitation is that this paper is still making classifications and doing analysis on isolated sections with $3.75$s of a consistent movement. To allow continuous control of a BCI we need to have predictions at every timestamp as proposed by \cite{blankertz2007non}. Solutions proposed to this continuous decoding as presented in \cite{tangermann2012review} may be a suitable direction.  %

The second limitation is that the classification performance of the models that we employed is very low for some subjects. For the subjects on which the highest performance was achieved an acceptable driving control would be possible. For the subjects with low performance (F1 $<0.6$) it is unlikely that the EEG-based BCI will allow for successful control. 

The problem of low-performing subjects may be addressed in future experiments by having participants use the EMG and EEG for control in a shared-control BCI \cite{mondini2020continuous, sburlea2021predicting}. By interleaving EMG predictions with EEG predictions and providing the subjects with feedback they can improve the separation of their EEG patterns, without reaching complete loss of control. Such a shared-control BCI should maximise the amount of EEG-based predictions to allow for learning, while still maintaining sufficient control with EMG. 

\printbibliography

\end{document}